\documentclass[conference]{IEEEtran}
\IEEEoverridecommandlockouts

\usepackage{cite}
\usepackage{booktabs}
\usepackage{amsmath,amssymb,amsfonts}
\usepackage{algorithmic}
\usepackage{graphicx}
\usepackage{hyperref}
\usepackage{textcomp}
\usepackage{subcaption}
\usepackage{xcolor}
\usepackage{url}
\usepackage[T1]{fontenc}
\usepackage[utf8]{inputenc}
\def\BibTeX{{\rm B\kern-.05em{\sc i\kern-.025em b}\kern-.08em
    T\kern-.1667em\lower.7ex\hbox{E}\kern-.125emX}}

\newcommand{\mason}[1]{\textcolor{black}{#1}}

\begin{document}

\title{Subtractive Training for Music Stem Insertion Using Latent Diffusion Models
\thanks{This work was supported in part by the National Science Foundation (NSF) under Grants DMS-2134248; in part by the NSF CAREER Award under Grant CCF-2236829 and in part by the Office of Naval Research under Grant N00014-24-1-2164.}
}

\author{
\textbf{Ivan Villa-Renteria}$^{*,1}$\thanks{* Equal contribution.}
\quad
\textbf{Mason Long Wang}$^{*,1}$
\quad
\textbf{Zachary Shah}$^{*,1}$
\quad
\textbf{Zhe Li}$^2$
\\
\textbf{Soohyun Kim}$^1$
\quad
\textbf{Neelesh Ramachandran}$^1$
\quad
\textbf{Mert Pilanci}$^1$\\
$^1$Stanford University
$^2$Hong Kong Polytechnic University
\\
}



\maketitle

\begin{abstract}
We present Subtractive Training\footnote{\url{subtractivetraining.github.io}}, a simple and novel method for synthesizing individual musical instrument stems given other instruments as context. 
This method pairs a dataset of complete music mixes with 1) a variant of the dataset lacking a specific stem, and 2) LLM-generated instructions describing how the missing stem should be reintroduced. We then fine-tune a pre-trained text-to-audio diffusion model to generate the missing instrument stem, guided by both the existing stems and the text instruction. Our results demonstrate Subtractive Training's efficacy in creating authentic instrument stems that seamlessly blend with the existing tracks. We also show that we can use the text instruction to control the generation of the inserted stem in terms of rhythm, dynamics, and genre, allowing us to modify the style of a single instrument in a full song while keeping the remaining instruments the same. Lastly, we extend this technique to MIDI formats, successfully generating compatible bass, drum, and guitar parts for incomplete arrangements.
\end{abstract}


\section{Introduction}
While text-based models for generating fully-mixed music \cite{dhariwal2020jukebox, agostinelli2023musiclm, copet2024simple, huang2023noise2music, liu2023audioldm} offer high-level control, they lack temporal or melodic precision, limiting their utility for musicians wanting to expand existing ideas. Ideally, tools should build upon musicians' inputs by adding complementary waveforms to enrich the piece. \mason{We observe that} music consists of multiple `stems'—individual audio waveforms which, when combined, form a complete piece. Stems are interdependent, meaning \mason{that} any subset imposes constraints on the rest. \mason{Musicians can iteratively build cohesive songs, stem-by-stem.}

We aim to generate stems that complement existing music using text-to-audio diffusion models, framing the task as spectrogram editing. Given a musical spectrogram and an instruction, we generate a new spectrogram that adds the specified stem while maintaining coherence. 

Inspired by image-editing work \cite{brooks2023instructpix2pix}, we propose Subtractive Training for diffusion models. This involves training on a dataset of complete music mixes, paired with corresponding incomplete mixes (where one stem is removed) and edit instructions. \mason{The dataset is created} via music source separation tools and a combination of music captioning and large language models. We fine-tune a text-to-audio diffusion model using the complete mixes as targets, and the incomplete ones with \mason{editing} prompts as input.

Our contributions are threefold. First, we demonstrate that our method can generate realistic, context-aware drum accompaniments. Second, we show that by modifying the text instruction, we can control the re-insertion of stems, allowing \mason{stem-wise} changes in arrangement, timbre, and style while preserving the rest of the instruments. We validate that Subtractive Training works with symbolic music, using a pitch-roll-based diffusion model to add guitar, bass, and drum stems.

\section{Background}

\subsection{Text-Based Image Editing}
Our method is inspired by InstructPix2Pix \cite{brooks2023instructpix2pix}, an image editing procedure that trains a diffusion model to edit images based on text instructions. The procedure uses GPT-3.5 Turbo \cite{brown2020language} and Stable Diffusion \cite{rombach2022high} to generate a large dataset of image-editing examples on which a conditional diffusion model is trained. Our method generates a similar dataset of text-guided spectrogram editing examples, focusing on edits that insert stems.

Our task is similar to image inpainting \cite{bertalmio2000image, elharrouss2020image, xiang2023deep}, where the goal is to infill masked portions of an image. However, instead of training the model to \emph{infill} portions of an image that have been \emph{masked}, we train the model to \emph{add} audio stems that have been \emph{subtracted}. Thus, in contrast to training procedures that are `masked,' our method is `subtractive,' hence the name `Subtractive Training.' 

\subsection{Diffusion Models}
Diffusion models have emerged as a powerful class of generative models, originating in the domain of image generation \cite{ho2020denoising, song2020denoising, rombach2022high}. These models learn to generate samples from a data distribution by iteratively denoising a Gaussian noise signal, gradually refining it into something that represents a generated sample. Many diffusion models operate in a latent space, using an encoder-decoder framework. In this framework, a variational autoencoder (VAE) \cite{kingma2019introduction} is employed to extract latent vectors that represent the desired data (images or audio). The diffusion model is then trained to iteratively denoise Gaussian noise signals into latent vectors that can be decoded by the VAE's decoder to generate data samples.

\subsection{Controlled Music Generation}
Since WaveNet \cite{oord2016wavenet}, there has been a surge of generative music models. Some are instances of latent diffusion models as described above \cite{Forsgren_Martiros_2022,huang2023noise2music, chen2024musicldm}. Other models use sequence modeling on audio tokens using transformers \cite{agostinelli2023musiclm, copet2024simple, garcia2023vampnet}. In the latter case, the training objective is to predict \emph{masked} tokens, while our method relies on \emph{subtracted} audio stems, as a natural analog to the conception of music as a sum of individual stems. This `subtractive' approach distinguishes itself from some other music editing frameworks, some of which predict masked tokens in music language models \cite{lin2024arrange}. Unlike previous work\cite{zhang2024instruct, han2023instructme, tal2024joint, lin2023content, nistal2024diff}, our method also uses LLMs to adapt text-to-audio models to ingest edit instructions by transforming descriptions of tracks into edit instructions. 





\section{Method}
\vspace{-2pt}

Inspired by \cite{brooks2023instructpix2pix}, the goal of our method is to provide a pre-trained text-to-audio diffusion model with a dataset of text-guided stem insertion examples. Unlike \cite{brooks2023instructpix2pix}, which uses cross-attention weight controls \cite{hertz2022prompttopromptimageeditingcross} to generate image pairs, we use a music source separation model to generate \mason{paired training examples}.  For each pair of complete and stem-subtracted spectrograms, we also use a music captioning model and preserve \cite{brooks2023instructpix2pix}'s use of a large language model to generate a text instruction describing how the missing stem should be added to complete the spectrogram.
Then, we fine-tune a pre-trained text-to-audio diffusion model on the task of \mason{denoising} the full-mix spectrogram, given both the text instruction and the stem-subtracted spectrogram.

\subsection{Dataset Generation}
\label{sec:training_data}
Our training requires audio-audio-text triplets consisting of (1) a full-mix spectrogram, (2) a version where one stem is removed, and (3) text instructions describing how to reinsert the missing stem. We construct a novel dataset using MusicCaps \cite{doh2023lp}, MUSDB18 \cite{rafii2017musdb18}, and MagnaTagATune \cite{law2009evaluation}, combined with source separation and music captioning tools.

\subsubsection{Full-Mix Spectrograms} 
We generate full-mix spectrograms by segmenting audio from MagnaTagATune and MusicCaps into 5 s chunks. For MUSDB18, we combine all stems, segment into chunks, and compute magnitude spectrograms.

\subsubsection{Stem-subtracted Spectrograms} For MagnaTagATune and MusicCaps, we use Demucs \cite{defossez2019demucs} to remove the stem of interest. 
For MUSDB18, this removal is done by excluding the stem of interest before combining. We segment stem-subtracted audio into 5 s chunks and compute spectrograms.

\subsubsection{Edit Instructions}
We generate edit instructions using LP-MusicCaps captions and GPT-3.5 Turbo. The model produces instructions describing how to reinsert a missing stem using action words (e.g., ``add'' or ``insert''). This results in 83.5k examples, each with paired full-mix and stem-subtracted spectrograms, captions, and edit instructions, which serve as input for our diffusion model fine-tuning.

\subsection{Subtractive Training}

Building upon the idea of generating missing stems based on existing musical context and text instructions, we propose an approach called Subtractive Training, defined as follows:

Consider the joint distribution $p(\mathbf{x}, \mathbf{y})$, where $\mathbf{x}$ represents a complete data sample and $\mathbf{y}$ represents an associated label or condition. In our context, $\mathbf{x}$ is a full-mix audio spectrogram, and $\mathbf{y}$ is an edit instruction describing how to add a specific instrument stem to the mix.

We decompose $\mathbf{x}$ into two components: $\mathbf{x}_{partial}$ and $\mathbf{x}_{missing}$, such that $\mathbf{x} = f(\mathbf{x}_{partial}, \mathbf{x}_{missing})$, where $f$ is a function that combines the two components to reconstruct the complete data sample. In our case, $\mathbf{x}_{partial}$ represents the stem-subtracted spectrogram (e.g., a song with the drum stem removed), and $\mathbf{x}_{missing}$ represents the missing instrument stem (e.g., the drum stem).

Our goal is to learn the conditional distribution $p(\mathbf{x} | \mathbf{y}, \mathbf{x}_{partial})$, which corresponds to the probability of generating the full mix given the edit instruction $\mathbf{y}$ and the stem-subtracted spectrogram $\mathbf{x}_{partial}$.

Diffusion models are particularly well-suited for this task, as they learn to model the data distribution by iteratively denoising a Gaussian noise signal conditioned on the input data. In our case, the diffusion model learns to generate the full mix $\mathbf{x}$ by conditioning on both the edit instruction $\mathbf{y}$ and the stem-subtracted spectrogram $\mathbf{x}_{partial}$. By training the model to estimate the conditional distribution $p(\mathbf{x} | \mathbf{y}, \mathbf{x}_{partial})$, we enable it to generate the missing instrument stem that follows the given edit instruction alongside the provided audio context.

\subsection{Fine-tuning the Diffusion Model}
We fine-tune a pre-trained text-to-audio latent diffusion model's U-Net \cite{ronneberger2015u} with audio-audio-text triplet data. When fine-tuning, the stem-subtracted spectrogram $\mathbf{x}_{partial}$, full-mix spectrogram $\mathbf{x}$, and edit instruction $\mathbf{y}$ are encoded into latents. We sample noise from a random timestep and add it to the full-mix latent, which is concatenated with the stem-subtracted latent and passed through the U-Net. The U-Net predicts the noise added to the full-mix latent, optimized via MSE loss.



\section{Experiments}

\subsection{Experimental Setup}

\begin{figure}[t!]
  \centering
  \includegraphics[width=\linewidth,trim={0 10 0 0},clip]
  {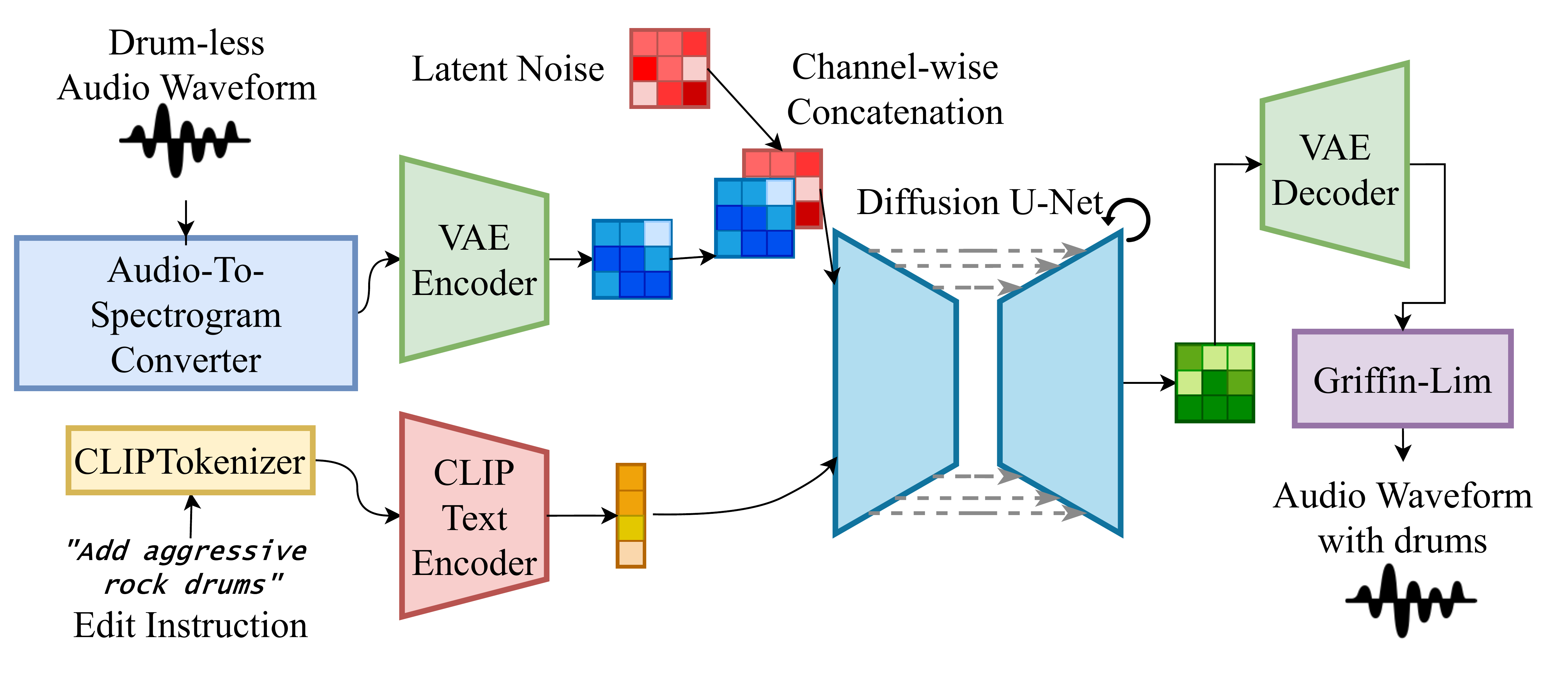}
  
  \caption{Latent Diffusion Model for Drum-Insertion.}
  \label{fig:model}

\end{figure}

\subsubsection{Model Architecture}
In our setup, we use Riffusion, a latent diffusion model that generates mel spectrograms conditioned on text prompts. Riffusion was created by fine-tuning the Stable Diffusion v1.5 checkpoint to operate on audio \mason{spectrograms}. The model accepts text prompts and 512x512 spectrograms as input and outputs 512x512 spectrograms. \mason{We modify this architecture to accept two channel inputs, corresponding to the noisy target spectrogram and the clean stem-subtracted spectrogram.} \mason{We show} the model architecture in Figure \ref{fig:model}.

As a baseline, we compare against SDEdit \cite{meng2021sdedit}, a diffusion-based style transfer method that is designed to edit images based on a given instruction, which we apply to Riffusion. 
We provide Riffusion with our stem-subtracted spectrogram and give it a text-conditioning signal instructing it to reinsert the missing stem. The SDEdit baseline adds a small amount of noise to the latent representation of the stem-subtracted spectrogram to enable editing, \mason{and denoises it according to the stem-insertion instruction. In contrast, our method adds noise to the full-mix spectrogram, while concatenating the clean stem-subtracted spectrogram as a condition. This avoids corrupting the information in the stem-subtracted spectrogram, which we want to preserve in the output.}

\subsubsection{Evaluation Dataset}

For evaluation, we create a separate test set using the MUSDB18 dataset \cite{rafii2017musdb18}. We extract 5 s clips from the MUSDB18 test split and perform the same stem subtraction, mel spectrogram computation, and edit instruction generation process as we did for the training data. Using the MUSDB18 test set for evaluation eliminates the effect of stem leakage on the generated outputs, since residuals from the drum track cannot be used to guide the drum insertion. 

    In total, the evaluation dataset contains 2,160 examples, each consisting of a 5 s full-mix clip, a corresponding stem-subtracted clip, and both a long-form text edit instruction and a shortened 5-word text caption. The short text captions are generated by prompting GPT-4 to summarize the full edit instructions, and are used as conditioning signals for the SDEdit baseline.

\subsubsection{Training Details}
We fine-tune the pre-trained Riffusion model on our dataset using the training procedure from InstructPix2Pix \cite{brooks2023instructpix2pix}. The weights of the VAE and the text encoder are frozen; only the UNet is updated. We train the model for 300k steps with a batch size of 4 on a single NVIDIA A10G GPU, which equates to roughly 15 epochs. We use the AdamW optimizer with $\beta_1=0.9, \beta_2=0.999$, weight decay of $0.02$, and learning rate of $10^{-4}$ with a cosine decay schedule and 500 warmup steps. We set the conditioning dropout probability to $0.05$ during training.

\subsubsection{Evaluation Metrics}
We evaluate our model and the SDEdit baseline using several metrics designed to assess the quality and diversity of the generated audio, following a similar procedure from \cite{liu2023audioldm}:

\begin{itemize}
\item \textbf{Fréchet Distance (FD)}: FD measures the similarity between generated and real audio features, extracted using a state-of-the-art audio classifier model called PANNs \cite{kong2020panns}. Lower FD indicates generated examples are more similar to real examples. 
\item \textbf{Fréchet Audio Distance (FAD)}: Our FAD metric is similar to FD, but uses features from a VGGish model \cite{hershey2017cnn} instead of PANNs. FAD may be less reliable than FD due to the potential limitations of the VGGish model.
\item \textbf{Kullback-Leibler Divergence (KLD)}: Measures the difference between the distributions of generated and real audio based on classifier predictions. Lower KLD suggests that the generated distribution is closer to the real data distribution \cite{vinay2022evaluating}. We compute KLD for each pair of generated and target examples and report the average.

\item \textbf{Inception Score (IS)}: Estimates the quality and diversity of generated examples based on the entropy of classifier predictions. Higher IS generally indicates better quality and diversity. \cite{vinay2022evaluating}
\end{itemize}

For the SDEdit baseline, we compare two variants using 20 and 50 denoising steps. Our model is evaluated using 20 denoising steps. Results on all metrics are shown in Table \ref{tab:results}.

\subsection{Quantitative Results}
\begin{table}[t!]
\centering
\begin{tabular}{lcccc}
\toprule
Model & FD$\downarrow$ & FAD $\downarrow$ & KLD $\downarrow$ & ISc $\uparrow$ \\ 
\midrule
SDEdit [20 Steps] & 33.25 & 4.40 & 3.79 & 1.35 \\
SDEdit [50 Steps] & 27.64 & 3.40 & 3.34 & 1.38 \\
S   .T. (Ours) & \textbf{5.55} & \textbf{0.62} & \textbf{1.10} & \textbf{1.41} \\ 
\bottomrule 
\end{tabular}
\caption{ Quantitative evaluation of the drums insertion task.}
\label{tab:results}
\end{table}

Table \ref{tab:results} presents the evaluation results comparing our method against the SDEdit baselines. Our model outperforms both SDEdit variants across all metrics, indicating that the outputs generated by our model are significantly closer to the target audio than those produced by SDEdit. Specifically, we observe a 22.09 decrease in Fréchet Distance and a 2.78 decrease in FAD compared to the best-performing SDEdit variant. Moreover, our method achieves a substantial 2.24 decrease in KLD and a modest increase in Inception Score from 1.38 to 1.41. 
These results show the effectiveness of Subtractive Training in generating high-quality, diverse drum tracks aligned with target audio.

\subsection{Qualitative Analysis}

\begin{figure}
\centering
\vspace{-10pt}
\begin{subfigure}[b]{0.32\columnwidth}
    \centering
    \includegraphics[width=\linewidth]{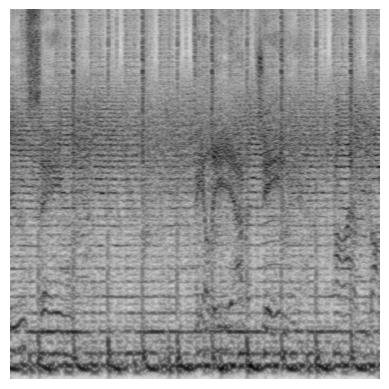}
    \caption{Full-mix}
    \label{fig:image1}
\end{subfigure}
\hfill
\begin{subfigure}[b]{0.32\columnwidth}
    \centering
    \includegraphics[width=\linewidth]{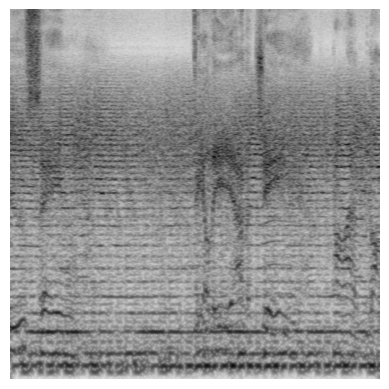}
    \caption{Stem-Subtracted}
    \label{fig:image2}
\end{subfigure}
\hfill
\begin{subfigure}[b]{0.32\columnwidth}
    \centering
    \includegraphics[width=\linewidth]{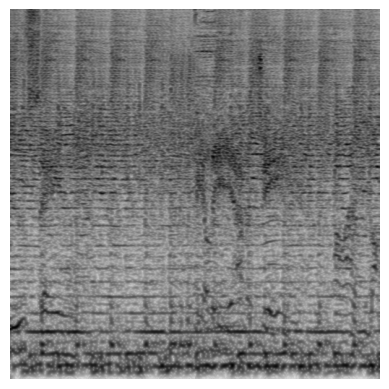}
    \caption{Generated}
    \label{fig:image3}
\end{subfigure}
\vspace{-5pt}
\caption{Comparison of spectrograms before and after stem addition. The stem-subtracted spectrogram is an input to our model, while the generated spectrogram is the output from the edit instruction "Add rock-style drums."}
\label{fig:spectrograms}
\end{figure}

\begin{figure}
\centering
\vspace{-10pt}
\begin{subfigure}[b]{0.32\columnwidth}
    \centering
    \includegraphics[width=\linewidth]{figs/images/rock_drums_target.png}
    \caption{Full-mix}
    \label{fig:rock_drums_target}
\end{subfigure}
\hfill
\begin{subfigure}[b]{0.32\columnwidth}
    \centering
    \includegraphics[width=\linewidth]{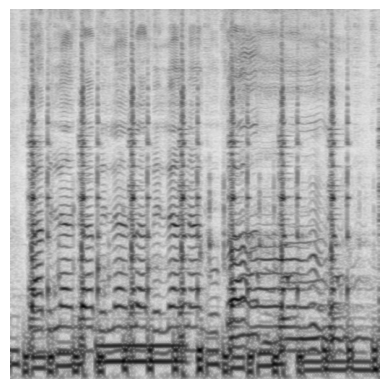}
    \caption{Stem-Subtracted}
    \label{fig:reggae_to_jazz_drums_background}
\end{subfigure}
\hfill
\begin{subfigure}[b]{0.32\columnwidth}
    \centering
    \includegraphics[width=\linewidth]{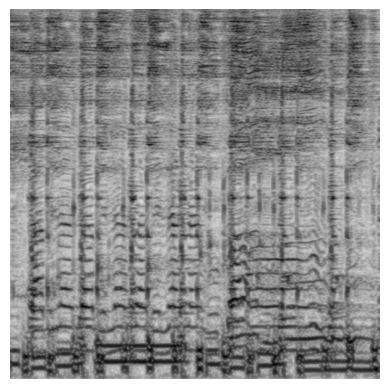}
    \caption{Generated}
    \label{fig:rjazz_drums_output}
\end{subfigure}
\vspace{-5pt}
\caption{Comparison of spectrograms before and after stem addition. The original genre of the song is reggae. Drums were inserted with the edit instruction ``add jazzy drums.''}
\label{fig:stylespectrograms}
\vspace{-5pt}
\end{figure}

We qualitatively examine the capabilities of our method with examples available on our website\footnote{subtractivetraining.github.io}. Figure \ref{fig:spectrograms} illustrates how our model integrates \mason{rock drums} into the input spectrogram while preserving the background content. The output does not replicate the pattern from the full mix, showing that diverse generations are possible without guidance from stem leakage. Additionally, our approach enables single-stem style transfer, \mason{through stem subtraction and subsequent addition. We can replace a reggae drum part with jazz-style drumming, shown in Figure \ref{fig:stylespectrograms}. The high frequencies in the generated spectrogram show the layering of rhythmically aligned jazz cymbals}. Our stem-specific editing and genre fusion enable new possibilities for musical creativity. Both examples (and more) can be listened to on our website.

\subsection{Listening Survey}
\begin{figure}
    \centering
    \hspace*{-0.07\linewidth}\includegraphics[width=1.1\linewidth]{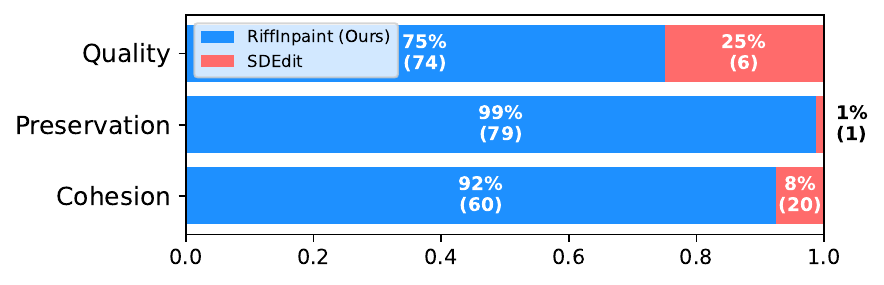}
    \caption{Subjective comparison of our method (RiffInpaint) to the SDEdit baseline. Win percentages and counts are shown for each method/criterion.}
    \label{fig:subjective}
    \vspace{-12pt}
\end{figure}

    We also conducted a subjective survey comparing our drum insertions to those of the SDEdit baseline. First, we play participants a randomly selected drumless song clip and inserted drums using our method and the SDEdit baseline. Then, participants select the preferred generation in terms of the sound quality (Quality), how well the original song clip is preserved (Preservation), and how well the generated drums fit with the original song (Cohesion). The results are shown in Fig.~\ref{fig:subjective}, and we include a link to the survey below\footnote{\url{https://mit.co1.qualtrics.com/jfe/form/SV_5yBc8RLuHwl93ro}}. We survey 10 users on 8 random examples from the test set. The survey demonstrates our method's superiority on all three criteria.

\subsection{Extension to MIDI}

To demonstrate the \mason{generality} of Subtractive Training, we extend it to symbolic music generation. We represent MIDI data as 3-channel piano roll images, where each channel corresponds to a specific instrument: drums, bass, or guitar. The piano roll values are binary, indicating presence/absence of a note at each time step and pitch.

We train three diffusion models, one for each instrument. For our architecture and training procedure, we use the binary-noise-based diffusion model described in \cite{atassi2023generating}. We use a large dataset of Guitar Pro tabs from DadaGP \cite{sarmento2021dadagp} to train our models, from which we transcribe 19,433 pitch-roll chunks.

\begin{figure}[t!]
  \centering
  \begin{subfigure}{\linewidth}
    \centering
    \includegraphics[width=\linewidth,trim=0.2cm 0cm 0cm 0cm, clip]{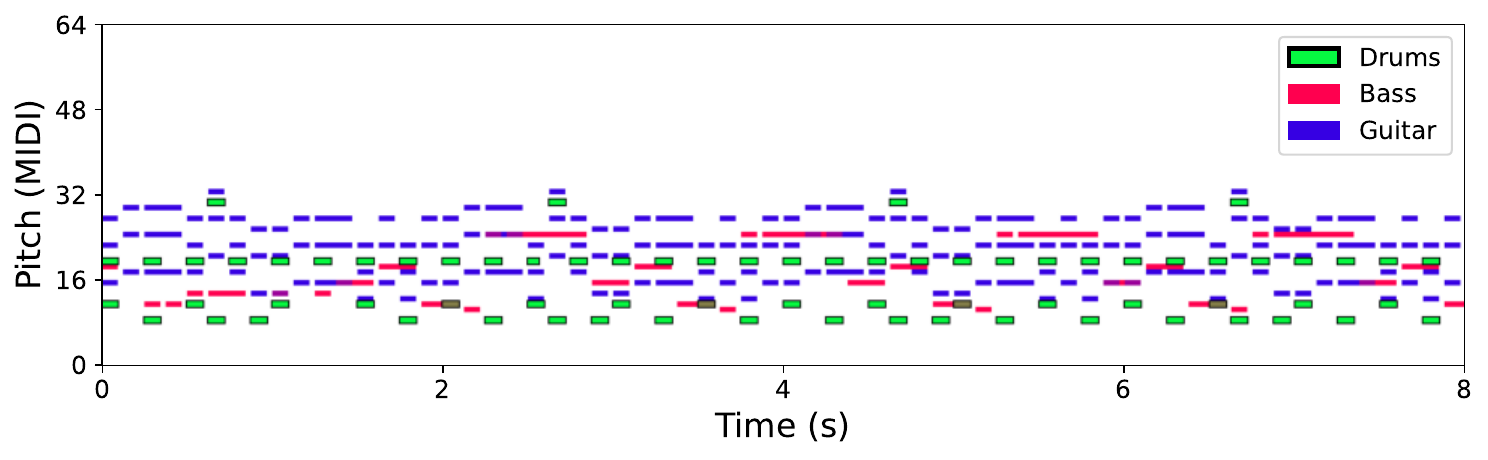}
    \vspace{-15pt}
    \caption{Drum Generation}

  \end{subfigure}

  \begin{subfigure}{\linewidth}
    \centering
    \includegraphics[width=\linewidth,trim=0.2cm 0cm 0cm 0cm, clip]{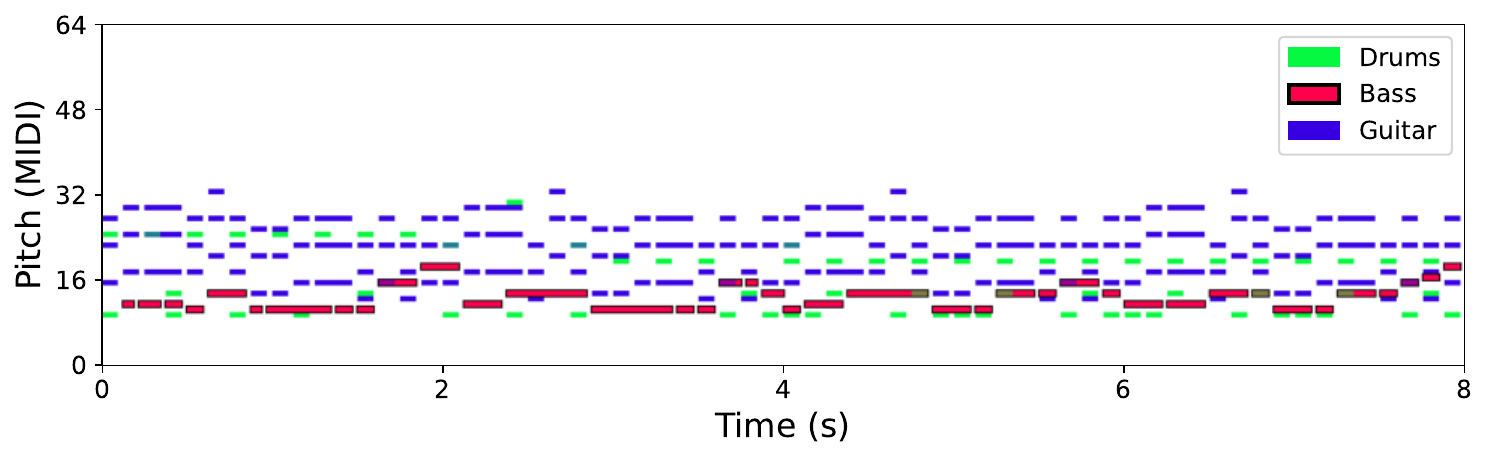}
    \vspace{-15pt}
    \caption{Bass Generation}

  \end{subfigure}



  \caption{Pitch rolls showing stem-generation results using two different diffusion models, each trained to output a given instrument. Notes corresponding to the generated instrument are outlined in black. Guitar generation examples can be found on the website.}
  \vspace{-12pt}
  \label{fig:midi}
\end{figure}

\mason{Like the other experiment, we concatenate noise to the clean context instruments and attempt to reconstruct the full mix.} Each model is fed a piano roll with two channels filled with the context instruments and the remaining channel initialized with noise. For example, the drum generator takes in piano rolls with the bass and guitar parts intact, but with the drum part replaced by noise. The diffusion process iteratively denoises the missing drum part conditioned on the bass and guitar parts. 

Figure \ref{fig:midi} shows generated results from held-out data, where we observe that notes generated with our model align well with the stems they are conditioned on. We provide several audio examples on our website and invite the reader to listen. Our model sometimes produces generations that are too sparse, which we remedy in practice by using rejection sampling.

\section{Discussion}

Future work can explore techniques to mitigate spectral leakage during stem separation. This would reduce the potential influence of residuals from the removed track on the generation. Our ideas include improving source separation, increasing the proportion of pre-stemmed data during training, and \mason{blurring} the spectrogram at problematic bands.


Subtractive Training shows promise in generating high-quality drum stems but faces challenges with high-frequency leakage and distinguishing synth sounds from artifacts in EDM tracks. Future work will also focus on synthetic data use, expanding to other stems, and enhancing edit instructions with advanced Music QA models.

\bibliographystyle{IEEEtran}
\bibliography{ref}

\end{document}